\documentclass[12pt]{article}
\usepackage{epsfig}
\usepackage{amssymb,amsmath}
\usepackage{setspace}
\pdfoutput=1
\usepackage{subfigure}
\usepackage{amssymb,amsmath}
\usepackage{graphicx}
\usepackage{color}
\usepackage{cancel}
\usepackage[colorlinks=true
,urlcolor=blue
,citecolor=blue
,linkcolor=blue
,pagecolor=blue
,linktocpage=true
,pdfproducer=medialab
]{hyperref}
\def\bi{\begin{itemize}}
\def\ei{\end{itemize}}

\def\to{\rightarrow}

\def\alt{\lesssim}
\def\agt{\gtrsim}

\usepackage{amsmath}

\newcommand{\bea}{\begin{eqnarray}}
\newcommand{\eea}{\end{eqnarray}}

\newcommand{\beq}{\begin{equation}}
\newcommand{\eeq}{\end{equation}}

 \makeatletter
\def\alt{\mathrel{\mathpalette\gl@align<}}
\def\agt{\mathrel{\mathpalette\gl@align>}}
\def\gl@align#1#2{\lower.6ex\vbox{\baselineskip\z@skip\lineskip\z@
\ialign{$\m@th#1\hfil##\hfil$\crcr#2\crcr\sim\crcr}}} \makeatother

\usepackage{longtable}

\begin{document}

\vspace*{1.0cm}

\begin{center}
\baselineskip 20pt {\Large\bf
Systematic Study of Diphoton Resonance \\ at 750 GeV from Sgoldstino
}
\vspace{1cm}

{\large
Ran Ding$^{a}$,
Yizhou Fan$^{b}$,
Li Huang$^{b}$,
Chuang Li$^{b}$, \\
Tianjun Li$^{b,c}$, 
Shabbar Raza$^{b}$,
Bin Zhu$^{d}$
} \vspace{.5cm}

{\baselineskip 20pt \it $^a$
Center for High-Energy
Physics, Peking University, Beijing, 100871, P. R. China
\\
{\it $^b$ 
State Key Laboratory of Theoretical Physics and Kavli Institute for Theoretical Physics China (KITPC),
Institute of Theoretical Physics, Chinese Academy of Sciences, Beijing 100190, P. R. China 
}
\\
{\it $^c$
School of Physical Electronics, University of Electronic Science and Technology of China,\\
Chengdu 610054, P. R. China 
}
\\
{\it $^d$
Institute of Physics, Chinese Academy of sciences, Beijing 100190, P. R. China
}
}

\vspace{.5cm}

\vspace{1.5cm} {\bf Abstract}
\end{center}
The ATLAS and CMS Collaborations of the Large Hadron Collider (LHC) have reported
an excess of events in diphoton channel with invariant mass of about 750 GeV.
With low energy supersymmetry breaking, we systematically consider
the sgoldstino scalar $S$ as the new resonance, which is a linear combination
of the CP-even scalar $s$ and CP-odd pseudoscalar $a$. Because we show that
$s$ and $a$ can be degenerated or have large mass splitting, we consider
two cases for all the following three scenarios: (1) Single resonance. $s$ is the 750 GeV resonance
decays to a pair of 1 GeV pseudoscalar $a$. With suitable decay length,
these two $a$ decay into collimated pair of photons which cannot be distinguished
at the LHC and may appear as diphotons instead of four photons.
(2) Twin resonances.
$m_{s}\approx m_{a}$ with a mass difference of about 40 GeV and both $s$ and $a$ decay
into diphoton pairs.
For productions, we consider three scenarios: (I) vector boson fusion; (II) gluon gluon fusion;
(III) $q{\bar q}$ pair production. In all these scenarios with two kinds of
resonances, we find the parameter space that satisfies the diphoton production cross section
from  3 to 13 ${\rm fb}$ and all the other experimental constraints. And we address
the decay width as well. In particular, in the third scenario, we  observe that
the production cross section is small but the decay width of $s$ or $a$ can be from 40 to 60 GeV.

\thispagestyle{empty}

\newpage

\addtocounter{page}{-1}

\baselineskip 18pt

\section{Introduction}
Recently, the
ATLAS~\cite{bib:ATLAS_diphoton} and CMS~\cite{bib:CMS_diphoton} Collaborations have announced an excess in diphoton channel with invariant mass of about 750 GeV at $\sqrt{s}=$ 13 TeV. Assuming  a narrow width resonance, the ATLAS Collaboration has reported a local $3.6\sigma$  excess at the diphoton invariant mass around 747~GeV with
an integrated luminosity of 3.2 ${\rm fb}^{-1}$.
But for a wider width resonance, the signal significance 
 increases to $3.9\sigma$ with a preferred width about 45~GeV.
 The CMS Collaboration, using 2.6 ${\rm fb}^{-1}$ of data, found a diphoton excess
with a local significance of $2.6\sigma$ at invariant mass around 760 GeV. The significance reduces to $2\sigma$, if the decay width around 45~GeV is assumed. 
The excesses in the cross sections can be roughly estimated as
$\sigma_{pp\to \gamma \gamma}^{13~ {\rm TeV}} \sim 3-13~{\rm fb}$~~\cite{bib:ATLAS_diphoton, bib:CMS_diphoton}.
 It should be noted that the
CMS Collaboration did search for diphoton resonance~\cite{Khachatryan:2015qba} at $\sqrt{s} =$ 8 TeV and observed a slight excess $\sim$ 2$\sigma$ at an invariant mass of about 750 GeV but on the other hand the ATLAS Collaboration
 did not go beyond the mass of 600 GeV for this channel~\cite{Aad:2014ioa}. This indicates that the present ATLAS and CMS
observations at $\sqrt{s} =$ 13 TeV LHC Run-II are consistent with their results at $\sqrt{s} =$ 8 TeV LHC Run-I for diphoton channel.

In this study we take these results optimistically and interpret the excess of diphoton events as a hint for 
new physics beyond the Standard Model (SM).
The observed resonance can be naively understood as a bosonic particle with mass $750$ GeV. This has triggered new studies in model building for both effective and renormalizable frameworks extensively~\cite{Mambrini:2015wyu, Chala:2015cev, Angelescu:2015uiz,Gupta:2015zzs,Altmannshofer:2015xfo, Ding:2015rxx, Knapen:2015dap, Agrawal:2015dbf, Dutta:2016jqn,
Petersson:2015mkr, Bellazzini:2015nxw, Demidov:2015zqn, Casas:2015blx, Bi:2015lcf}.

Supersymmetry is one of the most promising scenarios for new physics
beyond the SM. It provides unification of gauge couplings, solves gauge hierarchy problem, and provides dark matter candidate particles. 
It was pointed out in Refs.~\cite{Angelescu:2015uiz,Gupta:2015zzs,Altmannshofer:2015xfo} that in scenario like
2-Higgs Doublet Model (2HDM), including the Minimal Supersymmetric Standard Model (MSSM) and the Next-to-Minimal Supersymmetric
Standard Model (NMSSM), the branching ratio $Br(H/A\rightarrow \gamma\gamma)$ turns out to be very small ${\cal O}(10^{-6})$. It was further
noted that it remains small even in the extreme case of $\tan\beta\sim 1$ which is the lower limit required by the Renormalization Group Equation (RGE) running of Yukawa couplings. 
But if one relaxes symmetries like $R$-parity or introduce new interactions to the MSSM, one can address diphoton resonance
(see \cite{Ding:2015rxx, Dutta:2016jqn} and references therein).

In this paper, we consider the low energy supersymmetry breaking, {\it i.e.},
the $N=1$ supersymmetry is broken at low energy around TeV scale. So we have
a goldstino fermion $\tilde G$ and its superpartner sgoldstino $S=\frac{1}{\sqrt{2}}(s+ia)$
where $s$ and $a$ are a CP-even and CP-odd real scalars.
We systematically study the sgoldstino scalar $S$ as the new
resonance~\cite{Petersson:2015mkr, Bellazzini:2015nxw, Demidov:2015zqn, Casas:2015blx}.
We point out that $s$ and $a$ can be degenerated or have large mass splitting, so
we consider two cases for all the following three scenarios: 
(1) Single resonance. $s$ is the 750 GeV resonance
decays to a pair of 1 GeV pseudoscalar $a$. With proper decay length,
these two $a$ decay into collimated pair of photons which cannot be distinguished
at the LHC and may appear as diphotons instead of four photons
in the detector~\cite{Chala:2015cev, Knapen:2015dap, Agrawal:2015dbf, Bi:2015lcf}.
(2) Twin resonances.
$m_{s}\approx m_{a}$ with a mass difference around 40 GeV and both $s$ and $a$ decay
into diphoton pairs.
For productions, we consider three scenarios: (I) vector boson fusion (VBF); (II) gluon gluon fusion (gg-F);
(III) $q{\bar q}$ pair production. In the previous
papers~\cite{Petersson:2015mkr, Bellazzini:2015nxw, Demidov:2015zqn, Casas:2015blx},
only the second scenario has been
studied. In all these scenarios with two kinds of
resonances, we find the parameter space that satisfies the diphoton production cross section
from  3 to 13 ${\rm fb}$ and all the other experimental constraints. And we address
the decay width as well. In particular, in the third scenario, because the production cross section is small,
we can explain the large decay width around 45 GeV (from 40 to 60 GeV)
reported by the ATLAS Collaboration and
very easily accommodate the diphoton excess $\sigma^{13,18 \,{\rm TeV}}_{\gamma \gamma}$ simultaneously, and
all the current experimental constraints including dijet constraint can be satisfied as well.

This paper is organized as follows. We devote Section~\ref{sec:2} to describe our model
and provide a mechanism
to generate mass hierarchy in $m_{s}$ and $m_{a}$. In Section~\ref{sec:3}, we study the 750 GeV diphoton excess
in three scenarios with
two kinds of resonances in details.
 Conclusion and summary are given in Section~\ref{sec:4}.

\section{The Model Building}\label{sec:2}
We consider the low energy supersymmetry breaking model.  As a consequence of spontaneous global supersymmetry breaking,
there exists the Goldstino fermion goldistino and its superpartner sgoldistino ($S$), which is given as
$S=\frac{1}{\sqrt{2}}(s + ia)$, where $s$ and $a$ are the CP-even and CP-odd scalars respectively. In general, they can have different masses. In the later part of this section, we will present a possible mechanism to understand
the mass difference between $s$ and $a$, {\it i.e.}, the hierarchical mass
$m_{s}\gg m_{a}$ and the degenerated mass  $m_{s} \approx m_{a}$.
The Lagrangian of our model is given as 
\begin{align}
  -{\mathcal{L}}\supset \frac{m_{i}}{2\sqrt{2}F_{S}} F_{\mu\nu}^{i}
  \left(-sF^{\mu\nu i}+ia {\tilde F}^{\mu\nu i}\right)
+\left[\frac{S}{M}(y_{i}^{U}Q_{i}U_{i}^{c}H_{u}+ y_{i}^{D}Q_{i}D_{i}^{c}H_{d}) + A_S S^3 + {\rm H.C.} \right]~,~
\label{lagrangian}
\end{align}
\noindent where $m_i$  are gaugino masses from $F$-term of $S$, $\tilde X_{\mu\nu}=\frac{1}{2}\epsilon_{\mu\nu\alpha\beta}X^{\alpha\beta}$, and
we will discuss them more in the following.
Also, $y_{i}^{U,D}$ are the up-type and down-type Yukawa 
couplings, and $Q_{i},~U_{i}^{c},$ and $D_{i}^{c}$ are respectively
the left-handed up-type, right-handed up-type,
and right-handed down-type quraks, and $H_{u,d}$ are up-type and down-type Higgs doublets, respectively.

\subsection{The Mass Splitting between $s$ and $a$}\label{sec:2_1}

To have the mass splitting between $s$ and $a$, we consider
the following high-order K\"ahler potential
\begin{align}
\kappa =\kappa_{1} \frac{(S\bar S)^{2}}{M^2}+[\kappa_{2} \frac{S\bar S^{3}}{2M^2}+ \kappa_{2}^{*} \frac{S^{3}\bar S}{2M^2}]
\end{align}
One can write scalar potential as:
\begin{align}
V=\kappa_{1} \frac{|F_{S}|^{2}}{M^2}|S|^{2}+[\kappa_{2} \frac{|F_{S}|^{2}}{2M^2} \bar S^{2}+ \kappa_{2}^{*} \frac{|F_{S}|^{2}}{2M^2}{S}^{2}]
\label{scalar_pot}
\end{align}
Taking $\kappa_{2}$ real ($\kappa_{2}=\kappa_{2}^{*}$), $m_{S}^{2}=\frac{|F_{S}|^{2}}{M^{2}}$ and using $S=\frac{1}{\sqrt{2}}(s+ia)$, we can rewrite Eq.~(\ref{scalar_pot}) as follows
\begin{align}
V=(\kappa_{1}+\kappa_{2})\frac{m_{S}^2}{2}s^{2} + (\kappa_{1}-\kappa_{2})\frac{m_{S}^2}{2}a^{2}
\label{scalar_pot2}
\end{align}
Thus, we can have two simple cases for mass splitting: \\
{\bf Case (1):} Single resonance. 
When $\kappa_{1}\simeq \kappa_{2}$, we have $m_{s}\gg m_{a}$. We shall assume that 
$s$ is the 750 GeV resonance
decays to a pair of 1 GeV pseudoscalar $a$. With proper decay length,
these two $a$ decay into collimated pair of photons which cannot be distinguished
at the LHC and may appear as diphotons instead of four photons
in the detector~\cite{Chala:2015cev, Knapen:2015dap, Agrawal:2015dbf, Bi:2015lcf}. \\
{\bf Case (2):} Twin resonances.
When $\kappa_{1}\gg \kappa_{2}$, then $m_{s}\simeq m_{a}$.
We will consider $m_{s}\approx m_{a}$ with a mass difference around 40 GeV and
then both $s$ and $a$ decay into diphoton pairs.

\section{Productions and Decays of Sgolstino and the LHC Constrains}\label{sec:3}

In this section, we will study three scenarios for productions: (I) vector boson fusion (VBF);
(II) gluon gluon fusion (gg-F); (III) $q{\bar q}$ pair production. In the previous
papers~\cite{Petersson:2015mkr, Bellazzini:2015nxw, Demidov:2015zqn, Casas:2015blx},
only the second scenario has been considered. For decays into diphoton,
we consider the above two cases.
In our phenomenological studies, we employ FeynRules~\cite{Alloul:2013bka} to generate {\tt UFO} model file~\cite{Degrande:2011ua}, and {\tt MadGraph5\_aMC@NLO}~\cite{MG5} to calculate the production cross section
of $S$, and then check/verify our results with the package MSTW~\cite{MSTW}.

\subsection{Sgoldstino Production from Vector-Boson Fusion}\label{sec:3_1}

In this subsection we will study an effective production of sgoldstino $S$ through VBF and its subsequent decay into $W,Z$ and photons. For this purpose we use Eq.~(\ref{lagrangian}), and for simplicity we also assume that gluino mass $M_{3}$ and the Yukawa couplings $y_{i}^{U,D}$ are zero.
To give gluino mass, for simplicity,  we assume that gluino obtains the Dirac mass
via gauge mediation~\cite{Benakli:2010gi}. 
We see that $S$ couples to the electroweak gauge bosons in a direct way as below
\begin{align}
  - {\mathcal{L}}\supset& \frac{m_2}{\sqrt{2}F_{S}}W^{+\mu\nu}(-sW^{-}_{\mu\nu} +a \tilde W^{-}_{\mu\nu})+\frac{M_{ZZ}}{2\sqrt{2}F_{S}}Z^{\mu\nu}(-sZ_{\mu\nu} +a \tilde Z_{\mu\nu})
\nonumber \\
& + \frac{M_{Z\gamma}}{\sqrt{2}F_{S}}F^{\mu\nu}(-sZ_{\mu\nu} +a \tilde Z_{\mu\nu} )+ \frac{M_{\gamma\gamma}}{2\sqrt{2}F_{S}}F^{\mu\nu}(-sF_{\mu\nu} +a \tilde F_{\mu\nu}),
\label{lag_vbf}
\end{align}
\noindent where $M_{ZZ}=m_{1}\sin^{2}\theta_{W}+m_{2}\cos^{2}\theta_{W}$,
$M_{Z\gamma}=(m_{2}-m_{1})\cos\theta_{W}\sin\theta_{W}$, 
$M_{\gamma\gamma}=m_{1}\cos^{2}\theta_{W}+m_{2}\sin^{2}\theta_{W}$,
$m_{1,2}$ are gauginos masses corresponding to $U(1)_{Y}$ and $SU(2)_{L}$,
and $\theta_{W}$ is the weak mixing angle.

We employ a polynomial fitting function for VBF which is an approximate result from the vertex  functions
and scattering amplitude formula \cite{Azatov:2015oxa}, and
get cross section in terms of $m_{1,2}$ and $F_{S}$. We also vary $m_{1}$ and $m_{2}$. As we discussed above, in this scenario we will consider both Case (1) and Case (2) in details.

\begin{figure}[htp!]
\centering
\subfiguretopcaptrue
\subfigure{
\includegraphics[totalheight=6.5cm,width=8.cm]{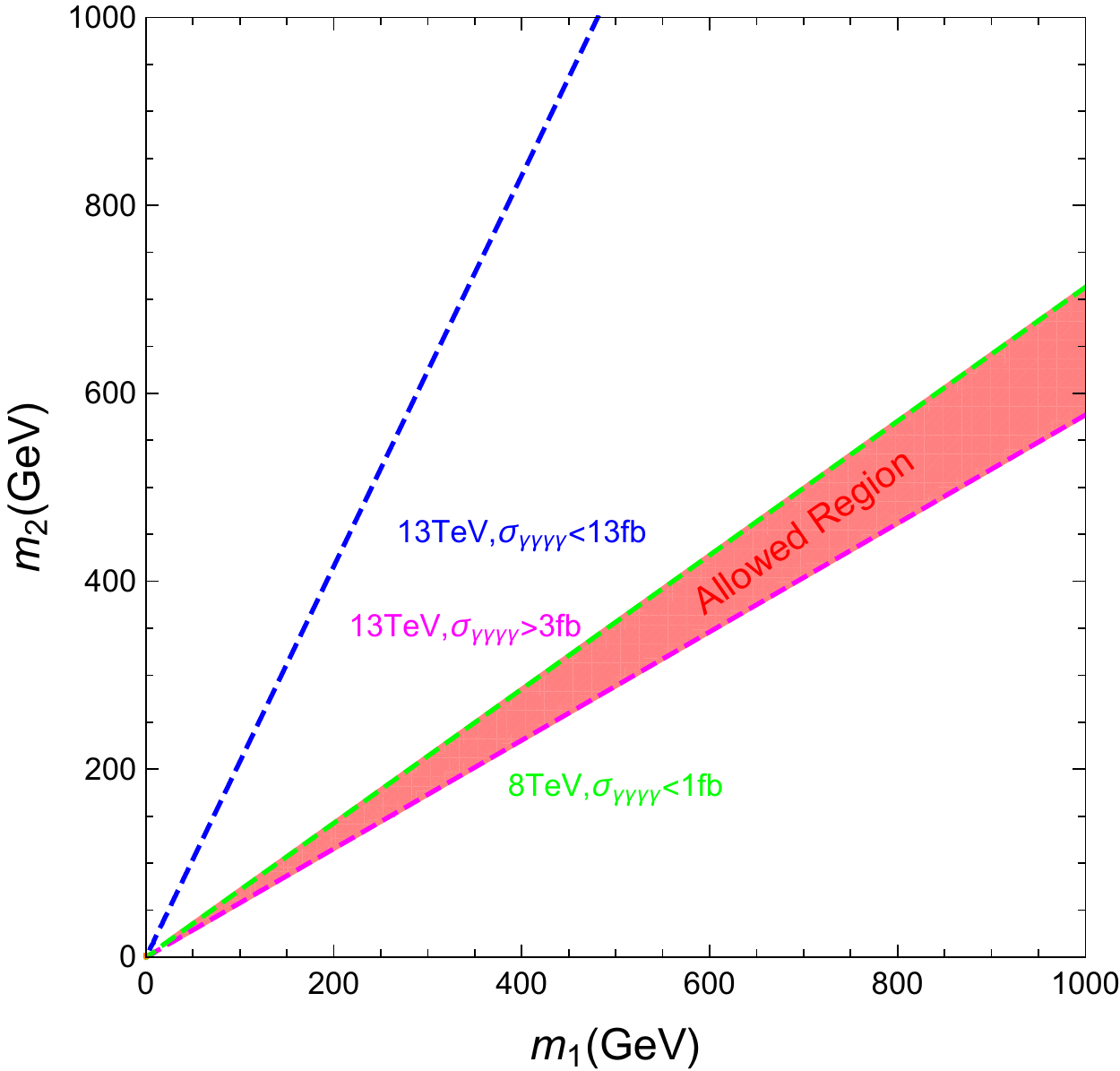}
}
\subfigure{
\includegraphics[totalheight=6.5cm,width=8.cm]{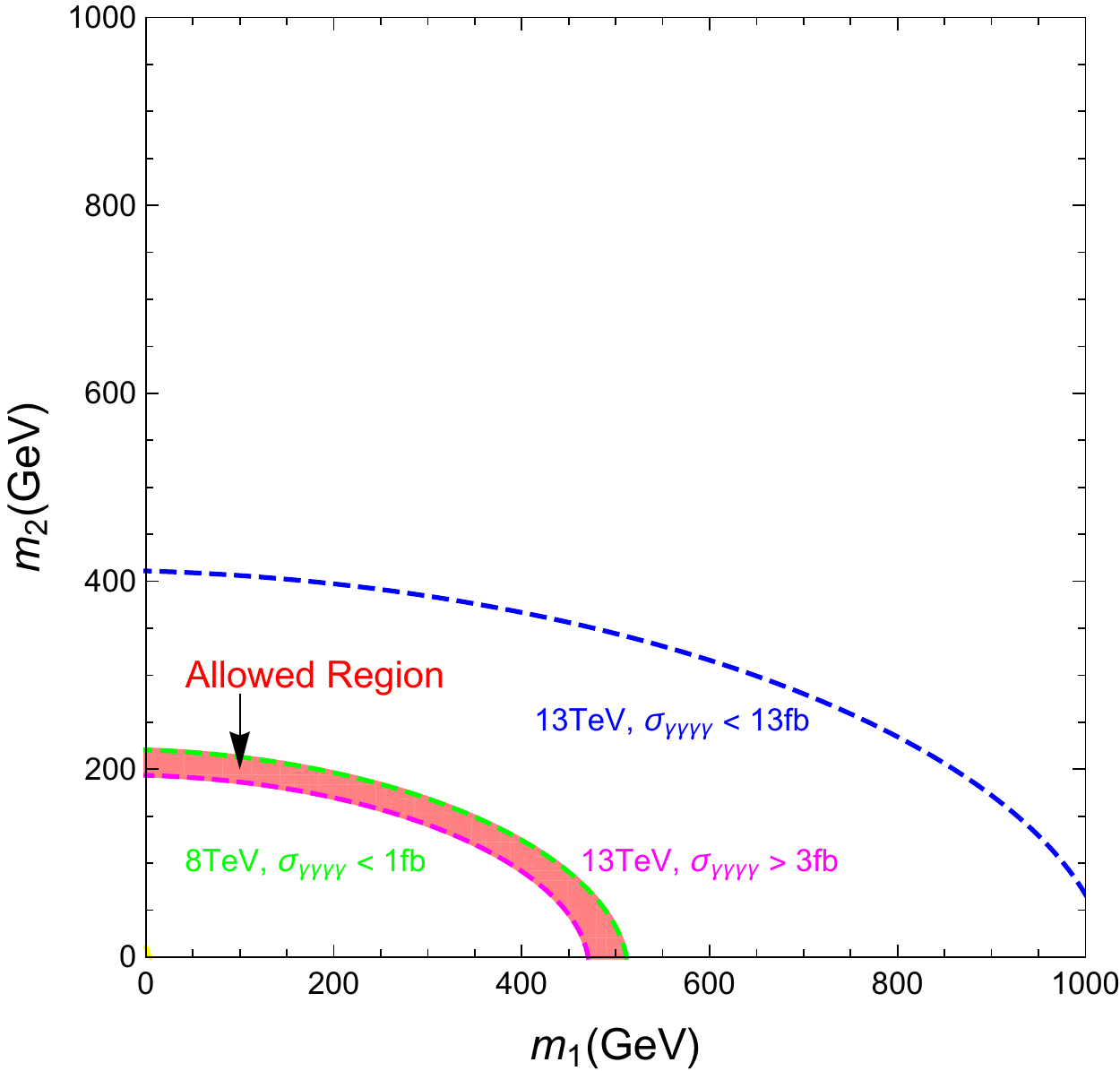}
}
\subfigure{
\includegraphics[totalheight=6.5cm,width=8.cm]{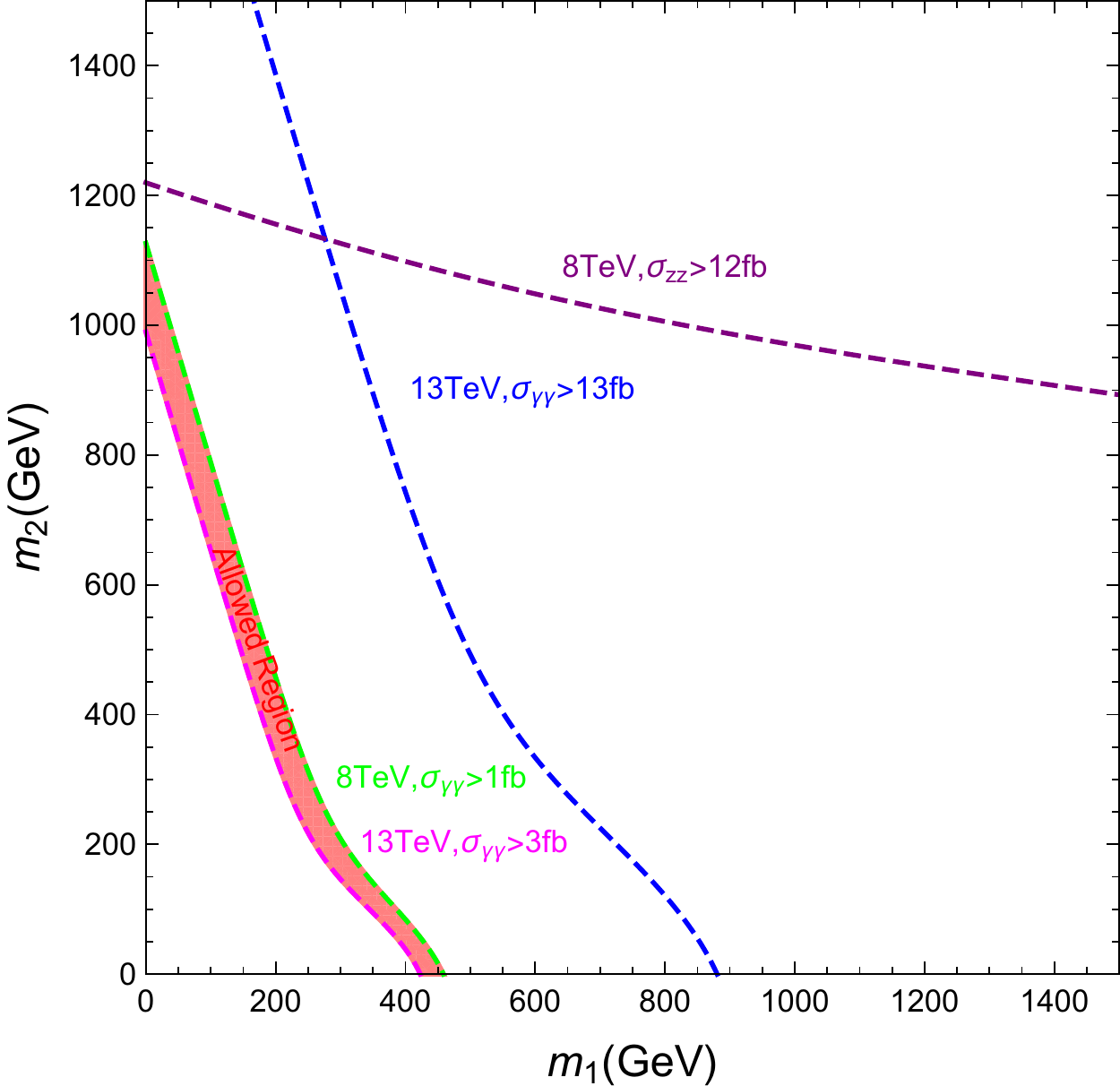}
}
\caption{Plots in $m_{1}-m_{2}$ plane. Top Left and right panels for $m_{s}\gg m_{a}$. Bottom panel for $m_{s}\approx m_{a}$.  }
\label{vbf_fusion}
\end{figure}

For Case (1) with $m_{s}\gg m_{a}$, we assume that the CP-even scalar $s$ of mass 750 GeV is produced
through VBF and then decays into a pair of pseudoscalar $a$ of mass $m_{a}=$ 1 GeV.
It has been argued in Refs.~\cite{Chala:2015cev, Knapen:2015dap, Agrawal:2015dbf, Bi:2015lcf}, if each of the pseudoscalar $a$ are light, highly boosted and they are not very long lived, then it is possible that
they decay into collimated pairs of photons. Then in the detector these photons may be measured as two photon events rather than four photon 
events. To study the process of production and decay of sgoldstino into diphoton, $pp\rightarrow S \rightarrow \gamma\gamma$ through VBF, we have to consider the following production-decay channels
{\flushleft
\begin{itemize}
\item [(i)] $pp\rightarrow V^{*}V^{*}\rightarrow s \rightarrow aa \rightarrow \gamma\gamma\gamma\gamma$
\item [(ii)]$pp\rightarrow V^{*}V^{*}\rightarrow s \rightarrow \gamma\gamma$
\item [(iii)]$pp\rightarrow V^{*}V^{*}\rightarrow s \rightarrow WW$
\item [(iv)]$pp\rightarrow V^{*}V^{*}\rightarrow s \rightarrow ZZ$
\item [(v)]$pp\rightarrow V^{*}V^{*}\rightarrow s \rightarrow Z\gamma$
\end{itemize}}
\noindent where $V^{*}V^{*} =WW,~ZZ,~{\rm and}~ Z\gamma$.
We require the following bounds on the parameter space:
\begin{align}
3 \,{\rm fb}\lesssim  &\,\sigma^{13 \,{\rm TeV}}_{ \gamma \gamma \gamma \gamma,\gamma \gamma}\lesssim 13 \,{\rm fb } ,
\label{con:1}
\end{align}
\begin{align}
                     &\,\sigma^{8 \,{\rm TeV}}_{\gamma\gamma \gamma \gamma,\gamma \gamma}\lesssim 1 \,{\rm fb } ,
\label{con:2}
\end{align}
\begin{align}
                     &\,\sigma^{8 \,{\rm TeV}}_{WW}\lesssim 30 \,{\rm fb } , 
\label{con:3}
\end{align}
\begin{align}
                     &\,  \sigma^{8 \,{\rm TeV}}_{ZZ}\lesssim 12 \,{\rm fb },
\label{con:4}
\end{align}
\noindent where $\sigma^{13,8 \,{\rm TeV}}_{XXXX}\equiv \sigma_{S}\times Br(S\rightarrow XXXX)$ and $\sigma^{13,8 \,{\rm TeV}}_{YY}\equiv \sigma_{S}\times Br(S\rightarrow YY)$.  
We make sure that $\Gamma_{s\rightarrow aa}$ is the dominant contribution. It should be noted that in this paper for $m_{s}\gg m_{a}$ we also make sure that $\Gamma_{s\rightarrow \gamma\gamma}$, $\Gamma_{s\rightarrow WW}$ and $\Gamma_{s\rightarrow ZZ}$ should be suppressed.  In addition to it, we also
demand that the total decay width of $s$ is $\Gamma_{s}=$ 40 GeV. Apart from these constraints,
another set of constraints come from the requirement to have collimated pair
of photons. In order to achieve this, as mentioned earlier, there are two conditions to take into account. First, the pseudoscalar $a$ should have a suitable decay length $l_{a}$ so it decays within the electromagnetic calorimeter (ECAL) of a detector. Second,  the opening angle $\alpha$ between the collimated photons in the Lab frame should be $\alpha \lesssim 4.6/\gamma$, here $\gamma$ is boost factor. It was shown in Refs.\cite{Chala:2015cev, Bi:2015lcf}, with this opening angle $\sim 90 \%$ of pseudoscalar $a$ can decay into collimated pair of photons. It should be noted that  the CMS ECAL has a resolution of $\Delta \eta \times \Delta \phi =0.0174 \times 0.0174$ and has radius R= 1.3 meters~\cite{Khachatryan:2015iwa} while for ATLAS has a resolution of $\Delta \eta \times \Delta \phi =0.025 \times 0.025$ and has radius R=1.5 meters~\cite{Aad:2009wy,Aad:2010sp}. For simplicity, if we assume that $a$ decays in the perpendicular
direction to the beam, we can have constrain on $R$ and $\Delta {\eta}$ \cite{Bi:2015lcf}:
\begin{align}
1 - \frac{375}{m_{a}\Gamma_{a\rightarrow \gamma\gamma} R } > 0, \\
\Delta \eta_{a} \approx \frac{4.6 m_{a}}{375}- \frac{4.6}{\Gamma_{a\rightarrow \gamma\gamma} R} \le \Delta \eta ,
\label{eta:bounds}
\end{align}
\noindent where $m_{a}$ is the mass of pseudoscalar, $\Gamma_{a\rightarrow \gamma\gamma}$ is the decay width of $a$ into a pair of gammas.
Here we note that the decay width of $\Gamma_{a\rightarrow \gamma\gamma}$ is given as
\begin{align}
\Gamma_{a\rightarrow \gamma\gamma}=\frac{(m_{1}\cos^{2}\theta_{W}+m_{2}\sin^{2}\theta_{W})^{2}m^3_{a}}{32\pi F_{S}^2}
\label{d-width}
\end{align}
\noindent while the decay length of $a$ can be written as
\begin{align}
 l_{a}=\frac{\gamma c}{\Gamma_{a\rightarrow \gamma\gamma}}.
\label{d-length}
\end{align}   
\noindent where $c$ is the speed of light. Using Eq.~(\ref{d-width}) in Eq.~(\ref{d-length}),
one can see $l_{a}\propto F_{S}^2$ so that $F_{S}$ cannot be arbitrary large, as we want to have $l_a$ within ECAL. We show our calculations in the top left panel of Fig.~\ref{vbf_fusion} in $m_{1}-m_{2}$ plane. In these calculations, we estimate $F_{S}=\frac{F_{max}(m_{1},m_{2})}{100}$ such that $l_{a}<R$. We apply the
constraints shown in Eqs.~(\ref{con:1})-(\ref{con:4}) and we take the CMS ECAL radius $R$=1.3 meters and $\Delta \eta$ = 0.017 to restrict $l_{a}$ and $\Delta \eta_{a}$. We show
$\sigma^{13 \,{\rm TeV}}_{\gamma \gamma \gamma \gamma }\lesssim 13 \,{\rm fb }$, $\sigma^{13 \,{\rm TeV}}_{\gamma \gamma \gamma \gamma}\gtrsim 3 \,{\rm fb }$ as blue and red dashed lines, while $\sigma^{8 \,{\rm TeV}}_{\gamma \gamma \gamma \gamma}\lesssim 1 \,{\rm fb }$. The allowed region is shown in red color. Here we want to make a comment that since we have considered $F_{S}$ as a linear function of $m_{1}$ and $m_{2}$, the couplings $\frac{m_{1}}{2\sqrt{2}F_{S}}\rightarrow {\rm constant}$ and $\frac{m_{2}}{2\sqrt{2}F_{S}}\rightarrow {\rm constant}$, when $m_{i}\rightarrow$ 0. This is why we see that the allowed regions is very small for smaller values of $m_{1}$ and $m_{2}$ but it gains width as $m_{1}$ and $m_{2}$ increases. Here we also note that $\sigma^{8 \,{\rm TeV}}_{WW,ZZ}$ is very small so the constraints $\sigma^{8 \,{\rm TeV}}_{WW}\lesssim 30 \,{\rm fb }$ and 
$\sigma^{8 \,{\rm TeV}}_{ZZ}\lesssim 12 \,{\rm fb }$ are not really effective here. 
For comparison, we also display our calculations for a fixed value of $F_{S}$, that is
$F_{S}=10^{6} \,{\rm GeV^{2}}$. In this case, since the couplings $\frac{m_{i}}{2\sqrt{2}F_{S}}\neq$ 0, we can expect curves in $m_{1}-m_{2}$ planes as can be seen in the top right panel. Color coding is the same as in the top left panel. Here we see that for $m_{2}=$ 0, $m_{1}\sim [450,500]$ GeV while $m_{2}\sim [190,225]$ for $m_{1}=$0.

In the Case (2) with $m_{s}\approx m_{a}$ that is the twin-resonance case,
we set $m_{s}=$ 750 GeV and $m_{a}=$ 710 GeV. So $s$ and $a$ can be produced via VBF and then decay into a pair of photons. In this way we can also explain the wide width of the observed resonance by the ATLAS collaboration. Remember that in twin-resonance case, there is no restriction on $l_{a}$ and $\Delta \eta_{s}$. Moreover, we also fix $F_{S}=10^{6} \,{\rm GeV^{2}}$. It should be noted that in this paper for $m_{s}\sim m_{a}$, we make sure to suppress $\Gamma_{s\rightarrow WW}$ and $\Gamma_{s\rightarrow ZZ}$. We show our results for this case in the bottom panel of Fig.~\ref{vbf_fusion}. Recall that in this case we have diphoton in the final state. For the constrains on diphoton final stated as indicated in Eqs.~(\ref{con:1})-(\ref{con:2}), we use the color coding is the same as in the top left panel. In this plot we also display $\sigma^{8 \,{\rm TeV}}_{ZZ}\lesssim 12 \,{\rm fb }$ in purple color.
We note that for $m_{1}\sim$ 0, the maximal allowed values of $m_{2}$ is about 1100 GeV. On the other hand, the maximal allowed value for $m_{1}$ is about 450 GeV for $m_{2}\sim$ 0.

\subsection{Sgoldstino Production from Gluon-Gluon Fusion}\label{sec:3_2}

Now we discuss the effective production of $S$ via gg-fusion (gg-F) and its subsequent decay into photons. Here again, for simplicity, we set the Yukawa couplings to be zero in Eq.~(\ref{lagrangian}). Then from the first part of Eq.~(\ref{lagrangian}), we see that in addition to couplings shown in Eq.~(\ref{lag_vbf}), there is
also a direct coupling of $S$ to gluon, for $m_{3}\neq$ 0, and can be given as:
\begin{align}
{\mathcal{L}}\supset& \frac{m_{3}}{2\sqrt{2}F_{S}} G^{a\mu\nu}(-sG^{a}_{\mu\nu}+
a \tilde G^{a}_{\mu\nu}). 
\label{eq:gluon}
\end{align}
 We vary $M_{1}$ and $M_{3}$ and keep
$M_{1}=M_{2}$ for simplicity. As we discussed above, in this scenario we will consider two possibilities that is
when $m_{s}\gg m_{a}$ and $m_{s}\approx m_{a}$. 
The processes $(i)-(v)$ indicated above can also be generated via gg-F. In addition to these process, we now also have, $pp\rightarrow gg \rightarrow S \rightarrow gg$.
This is why we also demand $\sigma^{8 \,{\rm TeV}}_{gg}\lesssim 30 \,{\rm pb }$ in addition to the constraints
shown in Eq.~(\ref{con:1})-(\ref{con:4}). We will use the narrow width approximation in this part of
our study. Besides suppressing $\Gamma_{s\rightarrow \gamma \gamma,WW,ZZ,gg}$, we neglect channels such as $\Gamma_{s\rightarrow gg}$ or $\Gamma_{s\rightarrow gggg}$ as they are too large to be fitted here.

\begin{figure}[htp!]
\centering
\subfiguretopcaptrue
\subfigure{
\includegraphics[totalheight=6.5cm,width=8.cm]{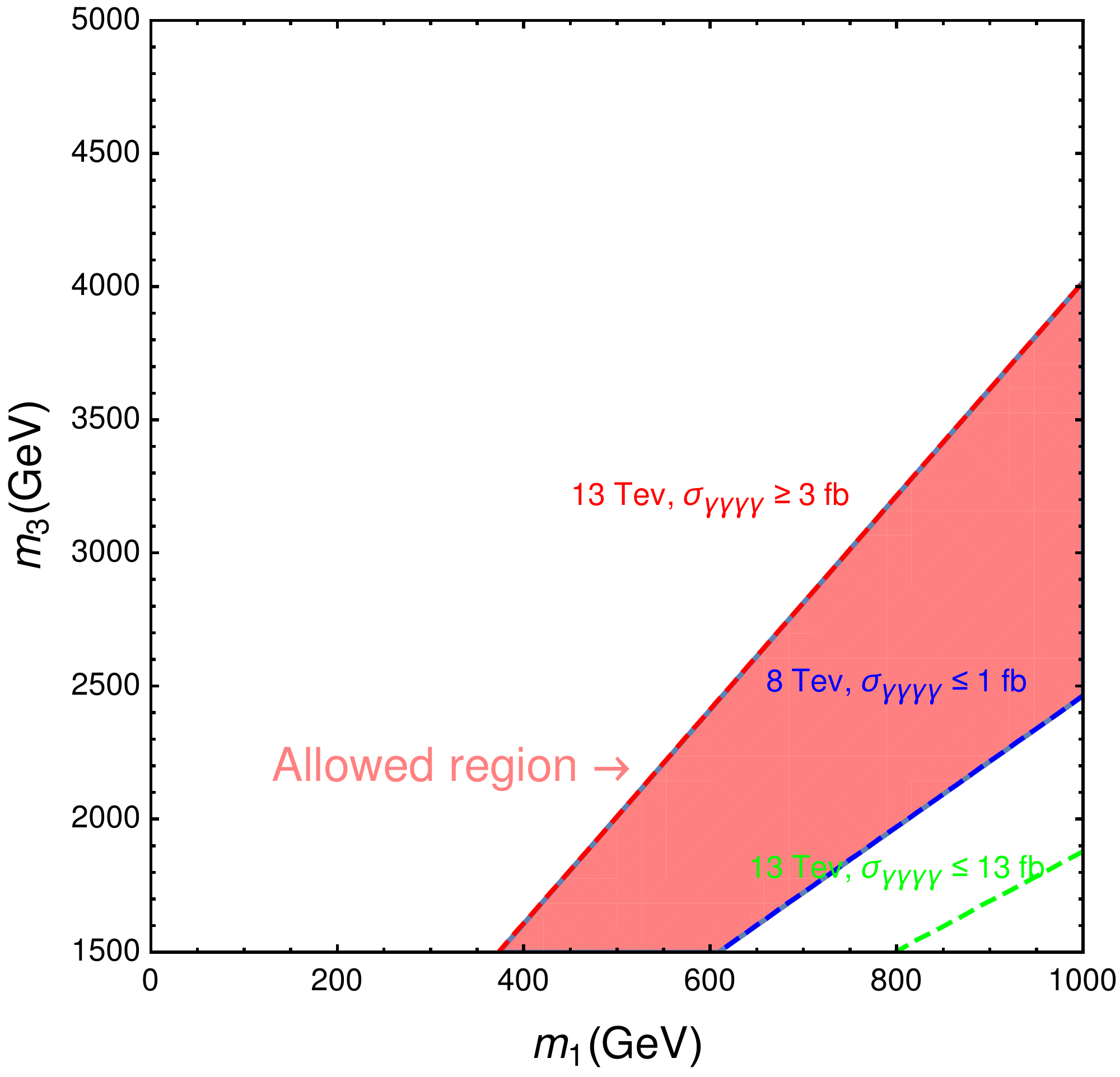}
}
\subfigure{
\includegraphics[totalheight=6.5cm,width=8.cm]{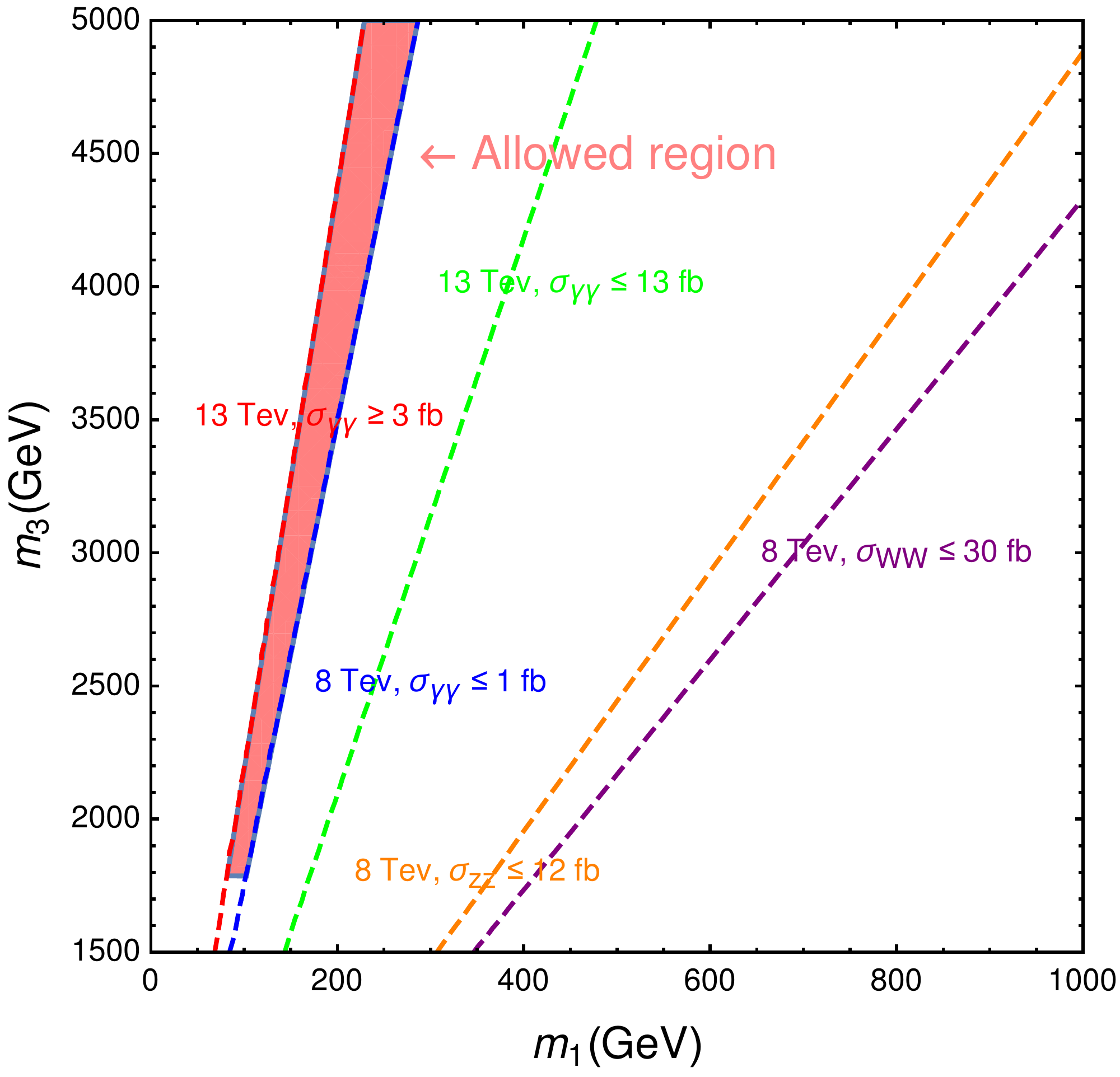}
}
\caption{Plots in $m_{1}-m_{3}$ plane. Left panel for $m_{s}\gg m_{a}$. Right panel for $m_{s}\approx m_{a}$.  }
\label{gg_fusion}
\end{figure}

We display our results in Fig.~\ref{gg_fusion} in $m_{1}-m_{3}$ plane. Plot in the left panel is for $m_{s}>> m_{a}$. We vary $F_{S}$ as we do in Section~\ref{sec:3_1}. For constraints $\sigma^{13, 8 \,{\rm TeV}}_{\gamma \gamma \gamma \gamma}$, the color coding is the same as in the left panel of Fig.~\ref{vbf_fusion}. The allowed region of parameter space is shown in red color. Here we see that for the allowed region, $m_{1}\sim[400,600]$ GeV while $m_{3}\sim[2500,4000]$ GeV. Plot in the right panel of Fig.~\ref{gg_fusion} represents the case $m_{s}\approx m_{a}$. Similar to Section~\ref{sec:3_1}, we set $m_{s} =$ 750 GeV and $m_{a}=$ 710 GeV.
As Eq.~(\ref{lagrangian}) shows, both of the particles can be produced via gg-F and decay into $ZZ,WW,Z\gamma,\gamma\gamma$, we have to consider all the constraints shown in Eqs.~(\ref{con:1})-(\ref{con:4})
and $\sigma^{8 \,{\rm TeV}}_{gg}\lesssim 30 \,{\rm pb }$. The color coding is the same as in the right panel of Fig.~\ref{vbf_fusion} but now we also display $\sigma^{8 \,{\rm TeV}}_{WW}\lesssim 30 \,{\rm fb }$ in orange color. The allowed parameter space is displayed in red color band. Here it can be seen that $m_{1}\sim [50, 250]$ GeV while $m_{3}\sim [1800, 5000]$ GeV.

\subsection{Sgoldstino from Quark-Antiquark Pair Production}\label{sec:3_3}

\begin{figure}[htp!]
\centering
\subfiguretopcaptrue
\subfigure{
\includegraphics[totalheight=6.5cm,width=8.cm]{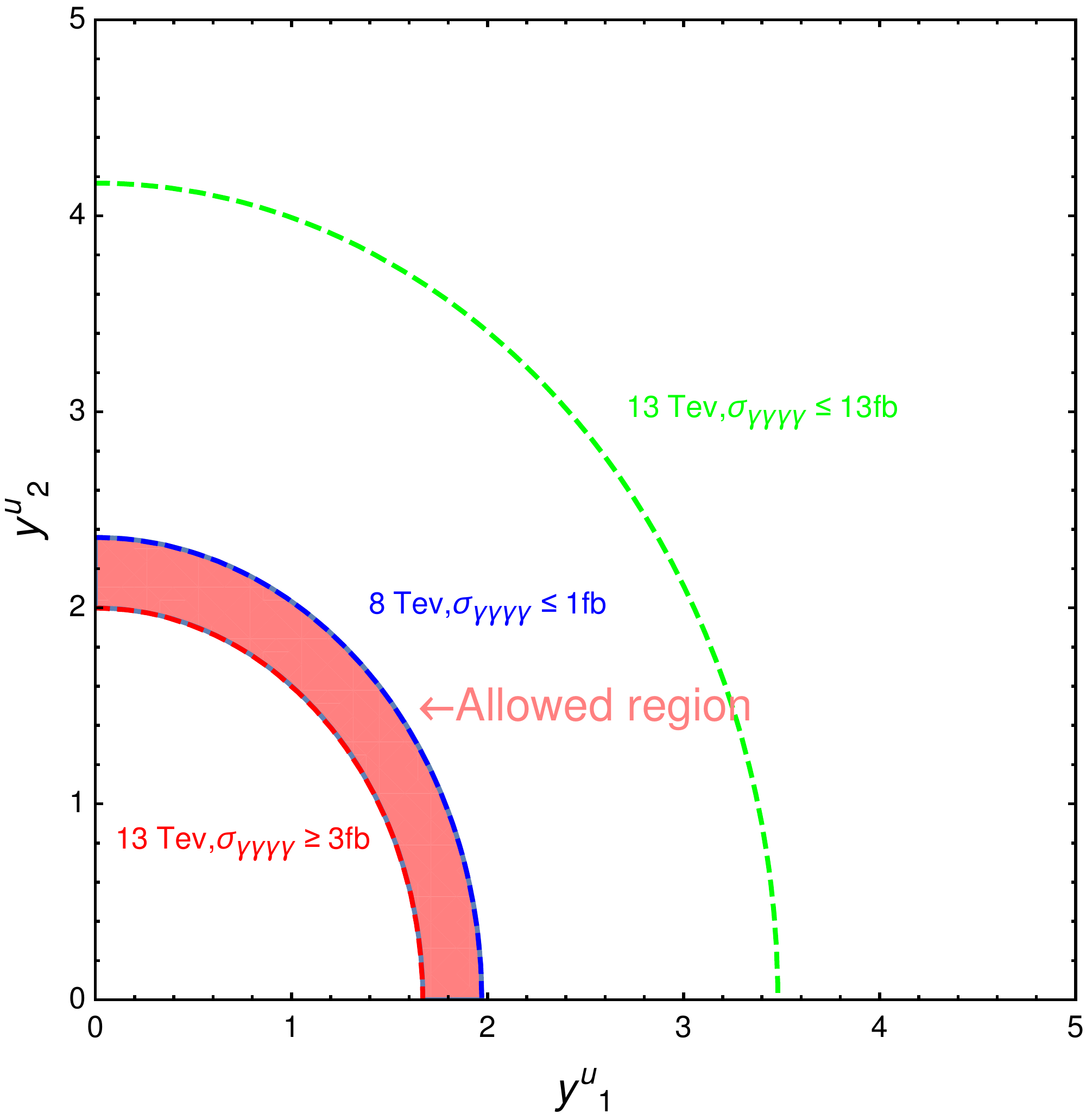}
}
\subfigure{
\includegraphics[totalheight=6.5cm,width=8.cm]{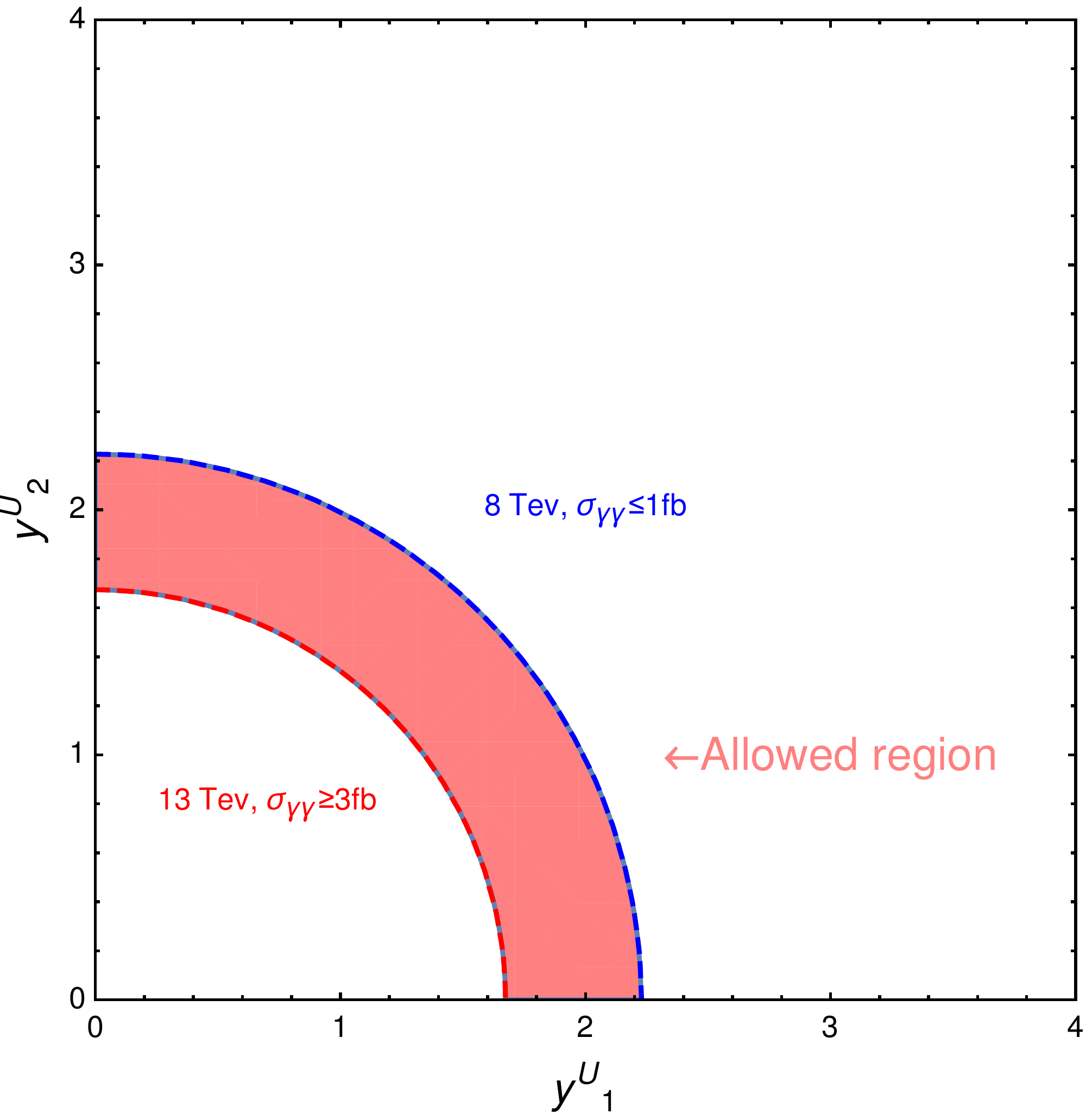}
}
\caption{Plots in $y_{1}^{U}-y_{2}^{U}$ plane. Left panel for $m_{s}\gg m_{a}$. Right panel for $m_{s}\approx m_{a}$.  }
\label{qq_fusion}
\end{figure}

In this subsection, we will study the production of sgoldstino $S$ from a quark-antiquark ($q\bar q$) pair. We find that in this case if we take $m_{1,2}=$ 500 and $F_{S}=5\times 10^{6} \,{\rm GeV^{2}}$ we can suppress VBF and by choosing $m_{3}=$ 0, gg-F can be avoided. And then we use second part of Eq.~(\ref{lagrangian}) for our study. For simplicity, we assume that the up-type and down-type Yukawa couplings of the first two families are same that is  $y^{U}_{1}=y^{D}_{1}$ and $y^{U}_{2}=y^{D}_{2}$. In this case too, as we considered earlier, we will assume $m_{s}\gg m_{a}$ and $m_{s}\approx m_{a}$ and use narrow width approximation. We find that the production cross section $\sigma_{q{\bar q}\rightarrow s}$ is small about 1500~${\rm fb}$ at $\sqrt{s}=$ 8 TeV. For the case $m_{s}\gg m_{a}$ we display our calculations in the left panel of Fig.~\ref{qq_fusion} in $y^{U}_{1}-y^{U}_{2}$ plane. The color coding is the same as in the left panel of Fig.~\ref{gg_fusion}. We see that for $y^{U}_{1}=$ 0, $y^{U}_{2}\sim [2,2.4]$, while for $y^{U}_{2}=$ 0, $y^{U}_{1}\sim [1.7,2]$.  We show our results for the case when $m_{s}\approx m_{a}$ in the right panel of Fig.~\ref{qq_fusion}. The color coding is the
same as in the right panel of Fig.~\ref{qq_fusion}. In this panel we observe that the allowed parameter band is somewhat larger as compared with the left panel. We notice that the maximal
allowed ranges for $y^{U}_{1}$ and $y^{U}_{2}$ are almost same as $\sim [1.7,2.2]$. Here we want to comment that since we can have large $y^{U}$, this implies large dijet cross section. But we find that in our case the dijet bound is very weak.

\begin{figure}[htp!]
\centering
\subfiguretopcaptrue
\subfigure{
\includegraphics[totalheight=6.5cm,width=8.cm]{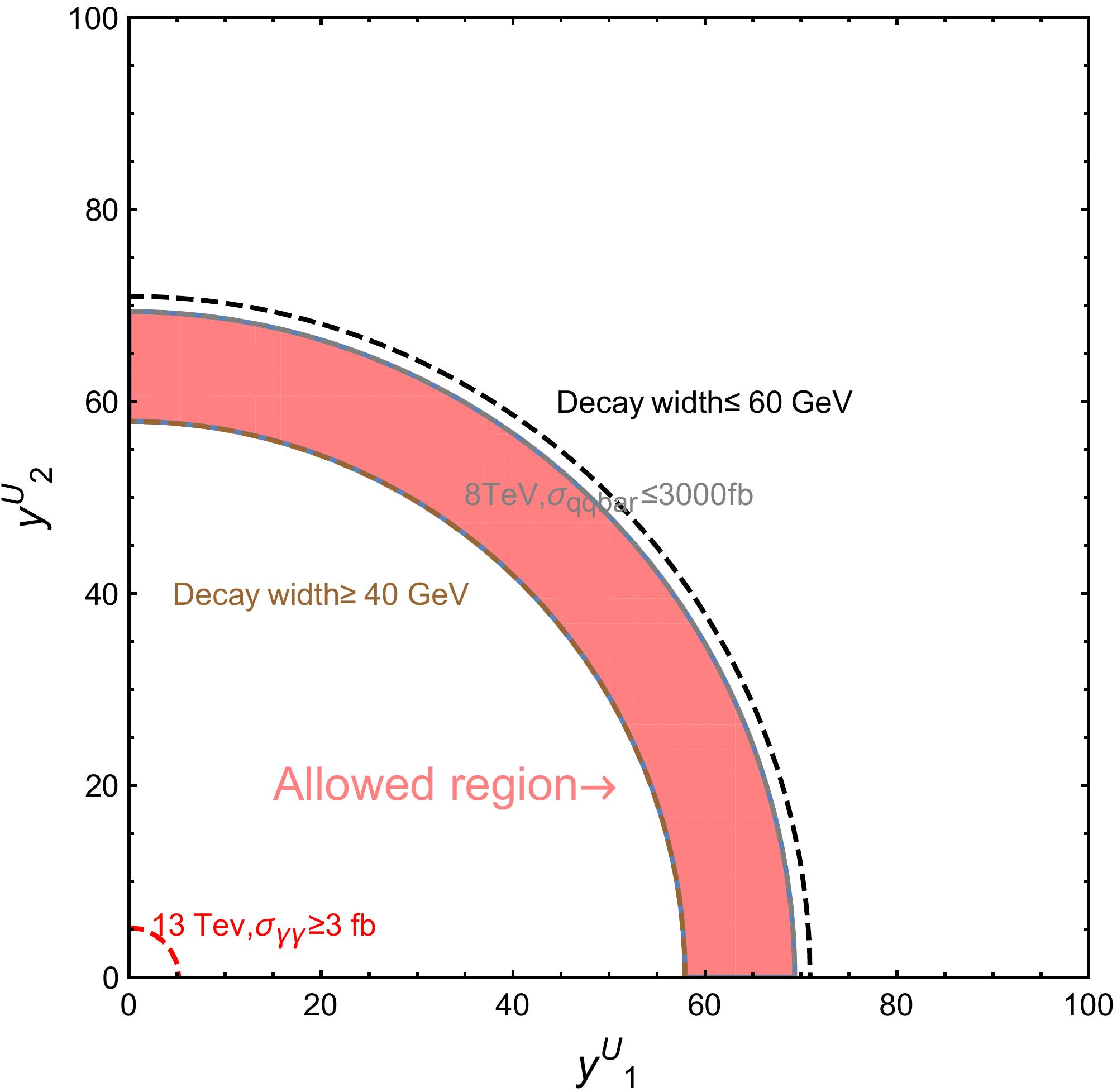}
}
\caption{Plots in $y_{1}^{U}-y_{2}^{U}$ plane.}
\label{qq_fusion_b}
\end{figure}

Now we discuss a very interesting scenario in which we can explain the large decay width of the resonance, in our case sgoldstino $S$ (either $s$ or $a$) from $q{\bar q}$ pair production. Let us
consider the cross section in the narrow width approximation as follows
\begin{align}
\sigma_{S} &= \sigma_{0} \Gamma(S\rightarrow q{\bar q}),
\end{align}
\noindent where  $\sigma_{0}$ is given by
 \begin{align}
 \sigma_{0}&=\frac{\pi^2}{8 M_{S}} \times \Big [\dfrac{1}{s}\dfrac{\partial \mathcal{L}_{q \bar q}}{\partial \tau} \Big ],  \nonumber \\
\dfrac{\partial \mathcal{L}_{q \bar q}}{\partial \tau} & = \int \limits_{0} dx_1 dx_2  f_{g}(x_1) f_g(x_2) \delta(x_1 x_2 - \tau), \nonumber
\end{align}
where $\tau = M^{2}_S/s$ and $\sqrt{s}=13$ TeV. We find that the value $\sigma_{0}$ is about 26$\rm{fb/GeV}$ and 110$\rm{fb/GeV}$ for 8 TeV and 13 TeV LHC respectively in contrast with 1400~$\rm{fb/GeV}$ for gg-F for 8 TeV LHC. This implies that the decay width $\Gamma(S\rightarrow q{\bar q})$ can be very large. In this way even if the cross section of $s$ or $a$ is not very large, bounds such as $\sigma^{13, 8 \,{\rm TeV}}_{\gamma \gamma}$
can be accommodated very easily. We presents our calculations in Fig.~\ref{qq_fusion_b}. Here we use $F_{S}=8\times 10^6 \,{\rm GeV^{2}}$. We display decay width less than 60 GeV and greater than 40 GeV in black and brown dashed curves. $\sigma^{8 \,{\rm TeV}}_{q \bar q}\le 3000 {\rm fb}$ is shown in blue dashed curve while $\sigma^{8 \,{\rm TeV}}_{\gamma \gamma}\le 3{\rm fb}$ is displayed in red dashed curve. Parameter space consistent will all the constrains is shown in red color. It can be seen from the plot that the maximum allowed ranges for $y^{U}_{1}$ and $y^{U}_{2}$ are almost same that is $\sim [57,70]$ .

\section{Summary and Conclusion}\label{sec:4}

The ATLAS and CMS Collaborations have reported
an excess of events in diphoton channel with invariant mass of about 750 GeV.
With low energy supersymmetry breaking, we systematically studied
the sgoldstino scalar $S$ as the new resonance, which is a linear combination
of the CP-even scalar $s$ and CP-odd pseudoscalar $a$.  We found that
$s$ and $a$ can be degenerated or have large mass splitting, so we considered
two cases for all the following three scenarios: (1) Single resonance. $s$ is the 750 GeV resonance
decays to a pair of 1 GeV pseudoscalar $a$. With suitable decay length,
these two $a$ decay into collimated pair of photons which cannot be distinguished
at the LHC and may appear as diphotons instead of four photons.
(2) Twin resonances.
$m_{s}\approx m_{a}$ with a mass difference of about 40 GeV and both $s$ and $a$ decay
into diphoton pairs.
For productions, we considered three scenarios: (I) vector boson fusion; (II) gluon gluon fusion;
(III) $q{\bar q}$ pair production. In the previous
literatures, only the second scenario has been studied.
In all these scenarios with two kinds of
resonances, we found the parameter space that satisfies the diphoton production cross section
from  3 to 13 ${\rm fb}$ and all the other experimental constraints. And we explained
the decay width as well. In particular, in the third scenario, we  observed that
the production cross section is small but the decay width of $s$ or $a$ can be from 40 to 60 GeV.

\section{Acknowledgements}
This research was supported in part by the Natural Science Foundation of China
under grant numbers 11135003, 11275246, 11475238 (TL).


\end{document}